\pdfoutput=1
\documentclass[
    ,final 
  ]
  {aipproc}

\layoutstyle{8x11double}

\begin{document}

\title{The Model Magnetic Configuration of the Extended Corona in the Solar Wind
Formation Region \footnote{\emph{Received 22 September 1998; accepted 12 April 1999;  Phys. Chem. Earth, Vol. 25, No. 1-2, pp. 113-116, 2000}}
}

\author{Igor Veselovsky}{
  address={Institute of Nuclear Physics, Moscow State University, Moscow 119899, Russia}
}
\author{Olga Panasenco}{
  address={Institute of Nuclear Physics, Moscow State University, Moscow 119899, Russia}
}

\begin{abstract}
The coupling between small and large scale
structures and processes on the Sun and in the heliosphere
is important in the relation to the global magnetic
configuration. Thin heliospheric current sheets
play the leading role in this respect.
The simple analytical model of the magnetic field configuration
is constructed as a superposition of the three
sources: 1) a point magnetic dipole in the center of the
Sun, 2) a thin ring current sheet with the azimuthal current
density $j_{\varphi}\sim r^{-3}$ near the equatorial plane and 3)
a magnetic quadrupole in the center of the Sun.
The model reproduces, in an asymptotically correct
manner, the known geometry of the field lines during
the declining phase and solar minimum years near the
Sun (the dipole term) as well as at large distances in
the domain of the superalfvenic solar wind in the heliosphere,
where the thin current sheet dominates and
$\vert B_{r}(\theta) \vert=const$ according to Ulysses observations
(Balogh et al., 1995; Smith et al., 1995). The model
with the axial quadrupole term is appropriate to describe
the North-South asymmetry of the field lines.
The model may be used as a reasonable analytical
interpolation between the both extreme asymptotic domains
(inside the region of the intermediate distances
$\sim (1-10)R_{\odot}$ ) when considering the problems of the solar
wind dynamics and cosmic ray propagation theories.
\end{abstract}

\maketitle

\section{1. Introduction}
The magnetic configuration of the extended solar corona
in the solar wind formation region plays an important
role when considering questions related to the analyses
of observed solar wind plasma characteristics and
energetic particles in the heliosphere. There are no direct
measurements of magnetic fields in these regions of
space, and available indirect data obtained by remote
sensing methods are scarce and sometimes contradictory
(see, e.g., Bird and Edenhofer, 1990; Suess, 1993).
Theoretical models, generally numerical ones are often
based on the use of potential, force-free or other MHDapproaches
without the sufficient attention to real electric
currents. Different theoretical approaches are described
in the literature (Schatten, 1971; Gleeson and
Axford, 1976; Wang, 1995; Zhao and Hoeksema, 1995).

The purpose of the present paper consists in an attempt
to fill the existing gap. A simple analytical model
is suggested taking into account qualitatively correctly
the main observational information about the structure
of the magnetic field in the extended solar corona and
the heliosphere during solar minimum years, when
the dipole component dominates. A model with the
quadrupole term is appropriate to describe the North-
South asymmetry.

The crucial point in choosing the model is the requirement
that it should reproduce asymptotically correct at
least the rough geometrical characteristics of the global
three-dimensional magnetic field known at present both
near the Sun and far from it.

\section{2. Simplest analytical model}

We consider the simplest model for solar minimum years,
which is represented as a superposition of fields from two
sources: a point magnetic dipole in the center of the Sun
and an extremely thin ring current sheet in the equatorial
plane with the surface current density  $j_{\varphi}\sim r^{-3}$.
The analysis made earlier (Veselovsky, 1994) showed
that it was a such concentrated current in the heliosphere
that created the magnetic field $\vert B_{r}(\theta) \vert=const$,
$B_{\theta}=0$. There emerged all supporting reasons in favor
of the present model for the description of the global
structure of the magnetic field due to the Ulysses measurements
that showed the $B_{r}(\theta)$ constancy in the broad
range of heliolatitudinal angles investigated up to $\pm80^\circ$.

One can neglect the solar rotation in the inner heliosphere,
where $\frac{\displaystyle\Omega r} {\displaystyle V} \ll1$. Just this domain we will
consider.

In the present model the magnetic field is expressed
by formulae
\begin{equation}
B_r= 2\mu r^{-3} \cos{\theta} +B_0 \left (\frac{\displaystyle r_0}{\displaystyle r}\right)^2 sign(\theta - \frac{\displaystyle \pi}{\displaystyle 2})
\end{equation}

\begin{equation}
B_{\theta}= \mu r^{-3} \sin{\theta} 
\end{equation}
where $\mu$ - the dipole moment, $|B_0| = \frac{\textstyle 2\pi r_0}{\textstyle c}|j_\varphi|$ -  value
of the magnetic field created by the surface current 
$j_{\varphi}~=~-~\biggl.\frac{\textstyle c}{\textstyle 2\pi}\frac{\textstyle {r_0}^2}{\textstyle r^3}\delta (\theta - \pi / 2) B_0$ (Veselovsky, 1994). The jump
of the field on the discontinuity equals $2B_0$.

It is convenient to use the physically plausible quantity
 $\Phi = B_0 {r_0}^2$, simply connected with the magnetic flux,
and to express through it the typical length $a = \biggl.\frac{\textstyle 2\mu}{\textstyle \Phi}$
arising in Eq. (1) as well as the typical field $B_a = \Phi a^{-2}$.
Then Eqs.~(1), (2) will become
\begin{equation}
B_r= B_a \rho^{-2} \biggl[\rho^{-1}\cos{\theta} + sign\bigl(\theta - \frac{\pi}{2}\bigr)\biggr]
\end{equation}
\begin{equation}
B_{\theta} = \frac{1}{2}B_a \rho^{-3} \sin {B} 
\end{equation}
where $\rho = \frac{\textstyle r}{\textstyle a}$ is the dimensionless length.

\section{3. Field structure}

The equation for field lines  $\frac{\textstyle dr}{\textstyle r\, d\theta}=\frac{\textstyle B_r}{\textstyle B_{\theta}}$ using Eq. (3) and
Eq. (4) gives
\begin{equation}
\frac{d\rho}{d\theta}=2\rho\bigl[\rho\,sign(\theta-\frac{\pi}{2})+\cos{\theta}\bigr](\sin{\theta})^{-1}
\end{equation}
Equation (5) is integrated in a straightforward way. Its
solution is
\begin{equation}
\rho = \sin^{2}{\theta}(C-2|\cos{\theta}|)^{-1}
\end{equation}
where C is an arbitrary constant of the integration,
which is the parameter of field lines.

Let us consider the obtained solution for two cases:
1) $\rho < 0\,(- \infty < a < 0)$; 2) $\rho > 0\,(\infty > a > 0)$, corresponding
to the existence of the thin current sheet with
one or another direction of the ring current. The special
case $|a| = \infty \, (\Phi = 0)$ corresponds to the field of the
point dipole in the absence of any current sheet.

1) Case $\rho < 0\,(a < 0)$.

In this case the dipole moment and the magnetic flux
of the ring current have identical signs. Zero points of
the field are absent.

\begin{figure}
\includegraphics[height=.26\textheight]{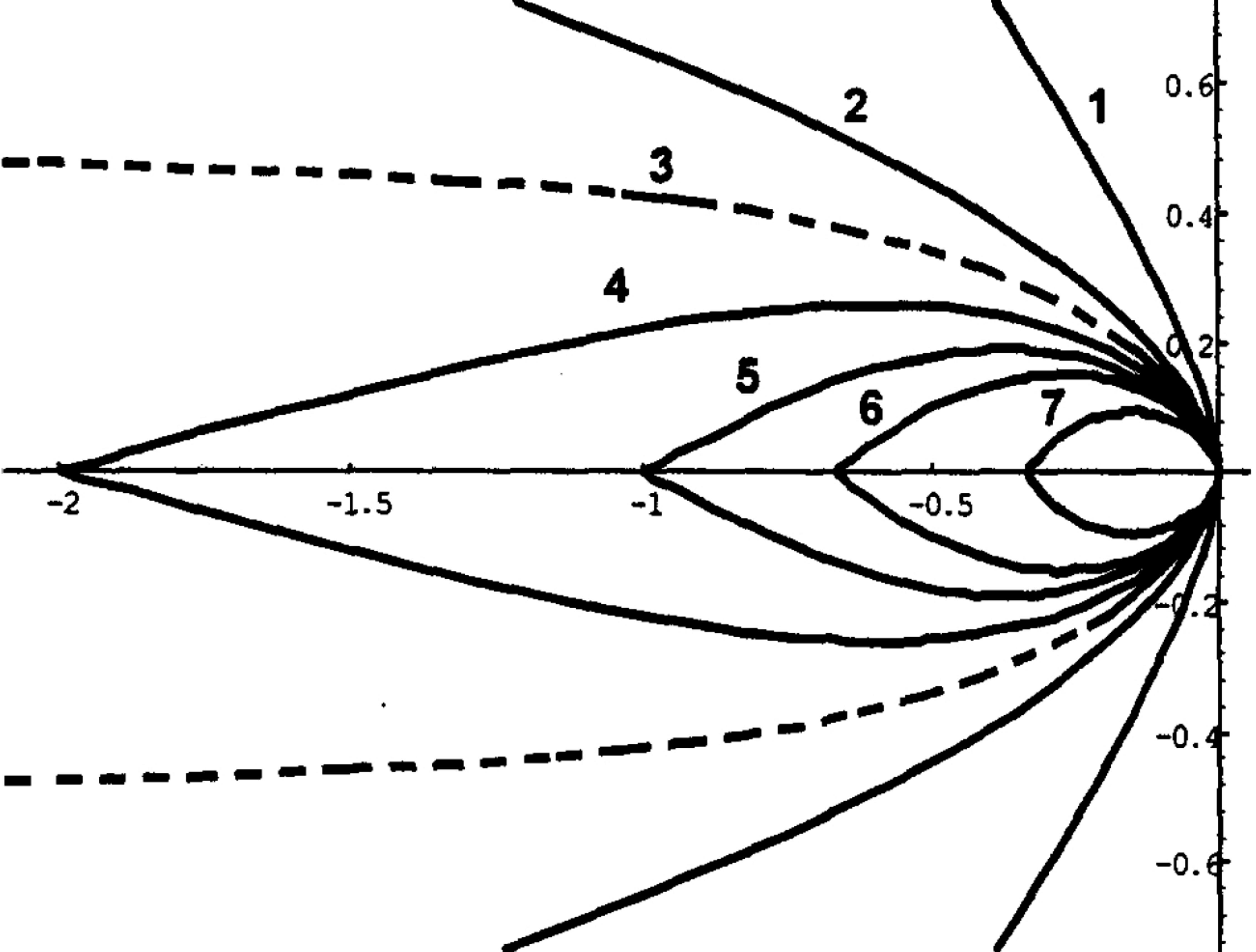}
\caption{The magnetic field lines calculated using Eq. 6 in the
case 1, when the point dipole and the magnetic flux of the thin
azimuthal ring current have the same signs (in the coordinate
system $(\rho, \theta)$): 1) C=1.5; 2) C=0.5; 3) C=0 i.e. the separator (dash
line); 4) C= =-0.5; 5) C=-1; 6) C=-1.5; 7) C=-3.}
\end{figure}

Any values of the parameter C are admissible in Eq.~(6),
$-\infty < C < + \infty$.

The closed ''dipole-like'' field lines $\rho(\theta)$ are obtained for
negative $C$ values, $C < 0$. These lines start and finish at
the coordinate system origin $\rho = 0$ going along $\theta \to 0, \pi $.
They asymptotically coincide with the lines of the dipole
field for $\rho \to 0$. The line with a given value $C$ crosses
the equator at the distance $\rho_e=\frac{\textstyle 1}{\textstyle C}$, that means $\frac{\textstyle a}{\textstyle C}$ in
the dimensional units. With $C < \frac{\textstyle 1}{\textstyle 2}$ the lines become
more tightened to the equatorial plane in comparison
with the lines of the dipole field, but with $C > \frac{\textstyle 1}{\textstyle 2}$ they
are less tightened. The fracture of the field lines takes
place when crossing the equator. The near equatorial
region $\theta \approx \pi /2$, occupied by the closed field lines, has an
asymptotic thickness 1 in dimensionless units at large
distances from the Sun.

The value $C=0$ corresponds to the separator
\begin{equation}
\rho = -\frac{1}{2}\sin{\theta}|\tan{\theta}|
\end{equation}
delimiting regions of closed and open field lines. In the
present model, this is a near equatorial boundary of two
polar coronal holes. Values $C > 0$ correspond to the
interiors of coronal holes, the domain of open field lines,
which are diverging as a fan from near polar regions at
the small distances ($\rho \to 0$ at $\theta \to 0, \,\pi$ according to
the dipole law). At large distances when $\rho\to\infty$ the
lines go out on to the radial asymptotics $\theta=\theta_{1,2}$ where
$\theta_{1,2}=\arccos{(\pm\, C/2)}$. The way out on to the asymptotics
 $\theta=\theta_{1,2}$ takes place from angles closer to the pole.
The last open line, which is closest to the equatorial
plane (the boundary of the polar coronal hole) passes
along the separator corresponding to the asymptotical
angle $\theta_{1,2}~=~\pi/2$. Figure 1 shows field lines calculated
according to Eq. (6) in this case.

2) Case $\rho > 0\,(a > 0)$.

The dipole moment and the magnetic flux of the ring
current are oppositely directed to each other. The field
has two zero points placed on the polar axis. The coordinates
of zero points in the coordinate system $(\rho, \theta)$ are
(1, 0) and (1, $\pi$). In the coordinate system $(r, \theta)$ these
coordinates are (a, 0), (a, $\pi$). All lines cross the equatorial
plane.
 \begin{figure}
  \includegraphics[height=.26\textheight]{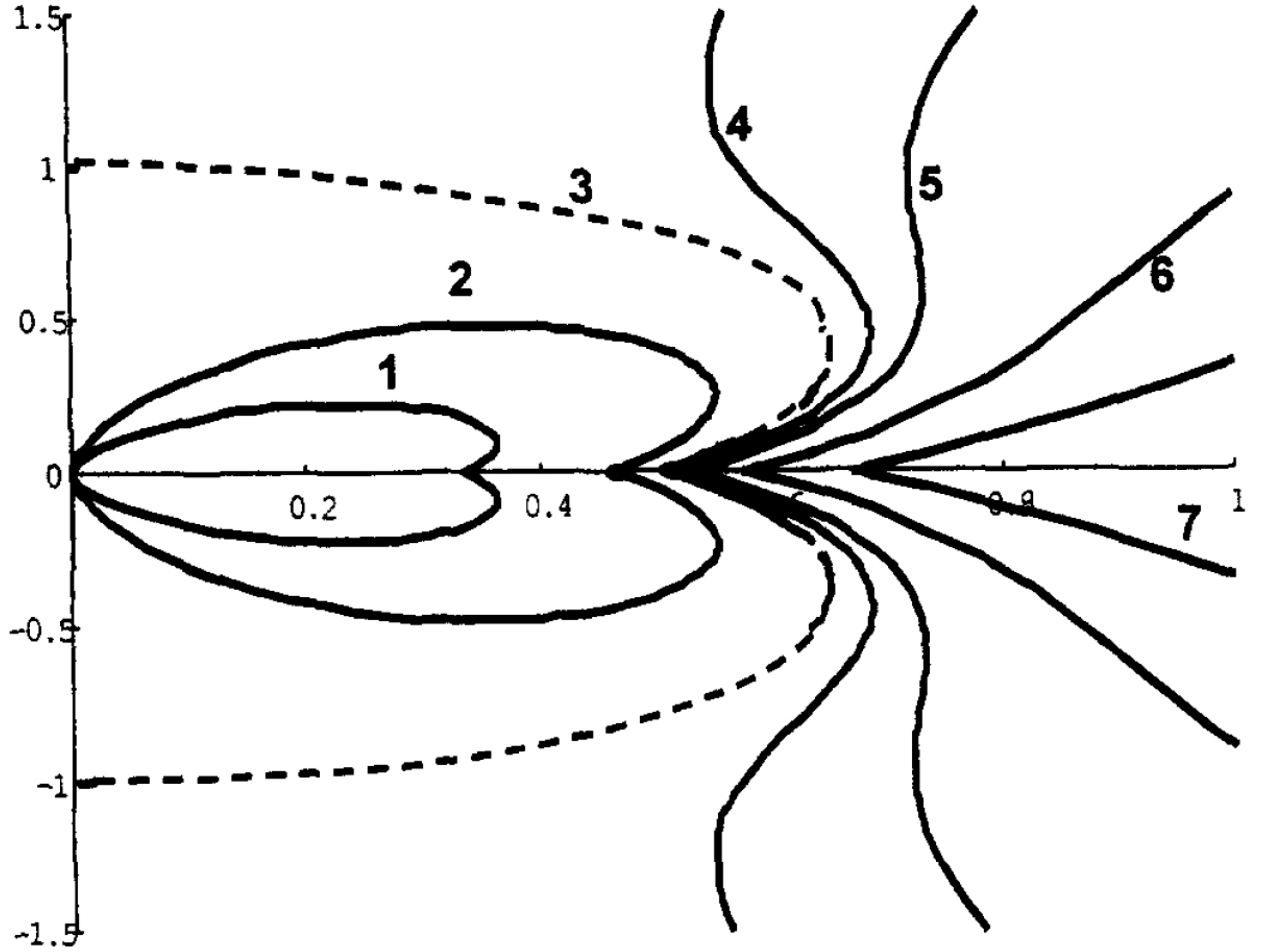}
\caption{The magnetic field lines calculated using Eq. 6 in the
case 2, when the point dipole is opposite to the magnetic flux of
the thin azimuthal ring current sheet (in the coordinate system
$(\rho, \theta)$): 1) C=3; 2) C=2.2; 3) C=2 i.e. the separator (dash line);
4) C=1.95; 5) C=1.9; 6) C=1.75; 7) C=1.5.}
\end{figure}
In this case only positive values of the parameter $C$ are admissible
in Eq. (6), $0 < C < +\infty$.

The closed ''dipole-like'' lines correspond to the values
$2 < C < +\infty$. They start and finish at the origin
of coordinates, crossing the equatorial plane at the distance
$\rho_e=\frac{\textstyle 1}{\textstyle C}$ (in the dimensional units $r_e=\frac{\textstyle a}{\textstyle C}$ ). The
value $C = 2$ corresponds to the separator, delimiting
the internal volume with closed lines from the external
volume with open lines. The equation for the separator
is
\begin{equation}
\rho = \frac{1}{2}\sin^{2}{\theta}(1-|\cos{\theta}|)^{-1}
\end{equation}
The separator is a surface of the oval shape around the
origin of coordinates and passes at the distance $r = a$
over the poles and crosses the equatorial plane at $r=\frac{\textstyle a}{\textstyle 2}$.

The region outside the separator is occupied by open
field lines. The values of the parameter $0 < C < 2$ correspond
to these lines. All such lines cross the equatorial
plane at the distance $\rho_e=\frac{\textstyle 1}{\textstyle C}$ and go to the infinity along
the radial asymptotical directions $\theta_{1,2}=\arccos{(\pm\, C/2)}$,
approaching these asymptotics at $r \to \infty$ from the side
of the equator. Figure 2 shows the calculated field lines
using Eq. (6) in this case.

\section{4. More complicated case}

Let us consider the potential of the magnetic field
\begin{equation}
\psi=\sum_{ l, m;\,l\ge1} {C_{l m}r^{-(l+1)}Y_{l m}(\theta, \phi)}+\frac{\Phi}{r} sign\,\theta(\phi)
\end{equation}
where $Y_{l m}(\theta, \phi)$ are the spherical harmonics, $\theta(\phi)$ is the
conic form of the heliospheric current sheet, $\Phi$ - the open
magnetic flux in the heliosphere.

In the simplest axially symmetric case with the only
equatorial current sheet $\theta(\phi) = const = \pi/2$ and $l~=~1,\;
m=0$, that is in agreement with observed structure of
the solar corona in the minimum years. The potential of
the magnetic field in the model with a quadrupole term
represented as
\begin{equation}
\psi=\frac{\mu}{r^2}Y_{10}+\frac{\kappa}{r^3}Y_{20}+\frac{\Phi}{r}sign\,\theta(\phi)
\end{equation}
where $\kappa$ is the quadrupole moment. The potentials of
the magnetic dipole $\psi_1$ and the quadrupole $\psi_2$ are expressed
by formulae
$$
\psi_1=\frac{\mu}{r^2}\frac{\cos{\theta}}{2}
$$
$$\psi_2=\frac{\kappa}{r^3}\,\frac{(-1+3\cos^2{\theta})}{4}
$$

In Figure 3 we show the global magnetic
patterns for dipolar and quadrupolar configurations on
the Sun together with the equatorial current sheets
$\theta(\phi) = const = \pi/2$ in our model. It is convenient to use
for the description of the Figure 3 following quantities: $\Phi$ -
the open magnetic flux, $\mu$ - the dipole moment, $\kappa$ - the
quadrupole moment, and to express through them the
typical independent lengths $a=-2\mu/\Phi$ and $b=-2\kappa/\mu$.
We consider two additional cases: 3) a < 0, b > 0; 4) a > 0, b > 0,
corresponding to the existence of the thin current sheet
with one or another direction of the ring current.

3) Case a < 0, b > 0.

In this case $\mu$ and $\Phi$ have identical signs, and $\kappa$ is opposite
to $\mu$ and $\Phi$.

4) Case a > 0, b > 0.

In this case $\mu$ and $\Phi$  are oppositely directed to each
other, and $\kappa$ is opposite to $\mu$. In Figure 3 the dash
line corresponds to the separator, delimiting regions of
closed and open field lines.

The quadrupole component decreases more rapidly
than the dipole component in the minimum years and at
a distance of $2R_\odot$ from the centre of the sun, the dipole component
clearly dominates. As an illustration, we have
chosen $|b/a|= |\kappa\Phi/\mu^2| = 0.25$ in our calculations.

\section{5. Electric circuit}

The existence of an equivalent electric circuit with radial
electric currents in the heliosphere was suggested by
H. Alfven (1977; 1981). Radial electric currents flowing
in the heliospheric current sheet of an order of several
$10^9 A$ are closed via volume electric currents inside the
polar regions.

\begin{figure}
\includegraphics[height=.3\textheight]{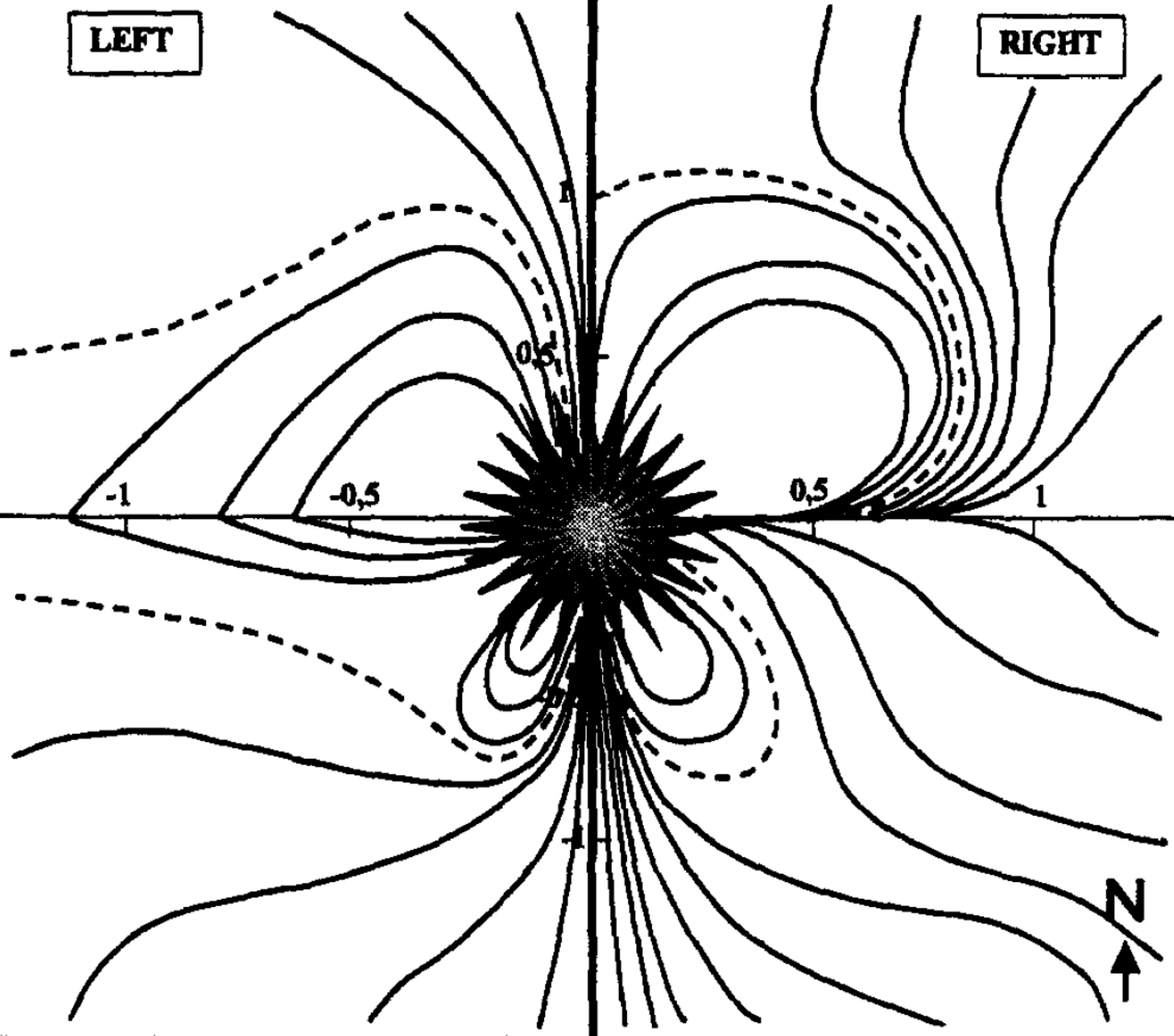}
\caption{The magnetic field lines for the case 3 are shown on the
left side. In this case the dipole moment $\mu$ and the magnetic flux
of the ring current $\Phi$ have identical signs, and the quadrupole
moment $\kappa$ is opposite to $\mu$ and $\Phi$. The magnetic lines for the
case 4 are shown on the right of the figure. In this case $\mu$ and $\Phi$
are oppositely directed each to other, and the quadrupole moment
$\kappa$ is opposite to $\mu$.}
\end{figure}

Azimuthal electric currents in the heliosphere are two
orders of magnitude stronger. They are forming a thin
ring current sheet around the Sun. The lines of this
surface current follow hyperbolic spirals orthogonal to
the Archimedian spirals of the magnetic field lines. Inhomogeneous
distributions of these surface currents are
manifested as strong heliospheric electrojets or partial
ring currents inside streamers (Veselovsky, 1994). Each
streamer in the corona contains a ribbon with a strong
horizontal current and closure currents flowing around
the streamer and partially diverted to the Sun along field
lines of the global magnetic field. Field-aligned currents
arise in this case in a manner analogous to the situation
with the partial ring current in the Earth's night-side
magnetosphere. The streamer belt consists, as a rule,
of several strong partial streamers and multiple weaker
entities. Hence, the system of inhomogeneous in-flowing
and out-flowing field-aligned electric currents appears in
both magnetic hemispheres, which should be partially
conjugated along global magnetic field lines.

It is important to note, that the main free magnetic
energy of the heliosphere is accumulated and available
for transformations inside the heliospheric current sheet
and its closure currents which are mainly field-aligned.

\section{Conclusions}
We have considered the simple analytical model of the
magnetic configuration of the extended solar corona in
the solar wind formation region. The model is represented
by the superposition of the known asymptotics
at the small and the large distances from the Sun.
During the years of low solar activity the overall geometry
near the Sun in the solar corona is dominated
by the large-scale field of the Sun. It is mainly represented
by the magnetic dipole slightly or moderately inclined
against the solar rotation axis, and the magnetic
field of the thin heliospheric current sheet situated near
the magnetic equator. The dipole inclination increases
with the solar cycle activity phase. Quadrupole and
higher harmonics and nonstationary perturbations are
especially important during higher solar activity years.
The strength of the heliospheric current sheet increases
and its geometry is more complicated during this period
of time. The overall oblate and curved shape of the
minimal corona and more spherical and radial maximal
structures are related to the evolution of the electric
current inside and outside the Sun with the solar cycle.

\begin{theacknowledgments}
This work was supported by the Federal Program \emph{''Astronomy''}
1.5.6.2 and the RFBF 98-02-17660 grant.
\end{theacknowledgments}

\end{document}